# Analysis of Profile and Morphology of Colloidal Deposits obtained from Evaporating Sessile Droplets

Laxman K. Malla[a], Rajneesh Bhardwaj[b, *], Adrian Neild[c]

[a] IITB-Monash Research Academy, Indian Institute of Technology Bombay, Mumbai, 400076, India.
[b] Department of Mechanical Engineering, Indian Institute of Technology Bombay, Mumbai, 400076 India.
[c] Department of Mechanical and Aerospace Engineering, Monash University, Melbourne, VIC 3800, Australia.
[*] Corresponding author (email: rajneesh.bhardwaj@iitb.ac.in)

*Abstract*

We experimentally investigate the profile and morphology of the ring-like deposits obtained after evaporation of a sessile water droplet containing polystyrene colloidal particles on a hydrophilic glass substrate. In particular, the coupled effect of particle size and concentration are studied. The deposits were qualitatively visualized under an optical microscope and profile of the ring was measured by an optical profilometer. The profile of the ring resembles a partial torus-like shape for all cases of particles size and concentration. The cracks on the surface of the ring were found to occur only at smaller particle size and larger concentration. We plot a regime map to classify three deposit types - discontinuous monolayer ring, continuous monolayer ring, and multiple layers ring - on particles concentration - particle size plane. Our data shows a possible existence of a critical concentration (particle size) for a given particle size (concentration) at which the monolayer ring forms. For the larger particle sizes, the immersion capillary forces between the particles dominate, aiding the formation of a monolayer ring of the particles. The relative mass of the particles accumulated in the ring is lesser in cases of the monolayer ring. We measure the width and height of the ring and show that they scale with particle concentration by a power law for the multiple layers ring. This scaling corroborates with an existing continuum based theoretical model. We briefly discuss the effect of the interaction of growing deposit with shrinking free surface on the ring dimensions and profile. The present results aid understanding of the ring formation process and will be useful in guiding the design of self-assemblies of the colloidal particles formed by the evaporating droplets.

## 1. Introduction

Understanding the mechanism of the formation of the deposit formed after evaporation of a sessile droplet containing colloidal particles (also known as "coffee-ring" problem [1]) is of much interest for technical applications such as inkjet printing, manufacturing of bioassays, designing droplet based biosensors and surface coating. This is a much-studied problem in colloids and interface science in the last two decades [2]. Recent studies [3,4] have demonstrated sorting of colloidal particles of different sizes suspended in such droplets, which has applications in designing complex self-assemblies. The well-established mechanism of the formation of a typical ring-like deposit is described as follows. In absence of convection outside the droplet, the evaporation is driven by the diffusion of liquid vapor into the ambient and the mass flux of the evaporation is largest near the contact line [1]. The contact line pins to the substrate due to contact angle hysteresis and an outward radial flow develops inside the droplet. This flow advects most of the suspended colloidal particles to the contact line, as shown schematically in Figure 1(a) and consequently, a ring-like deposit forms (Figure 1(b)).

Recent reviews by Routh [5] and Parsa et al. [6] identified a multitude of parameters, which could play a role in determining the pattern of the final deposit: Marangoni convection, pH of the suspension, particle size, particles shape, particle concentration, substrate temperature, substrate wettability, particles hydrophobicity, relative humidity, electrowetting etc. In particular, Marangoni convection has been shown to reverse the coffee-ring effect [7–10] and an inner deposit with a much smaller radius than the initial wetted radius is the final outcome instead of the ring. By lowering the pH [11] or by using ellipsoidal particles [12], or by applying electrowetting [13, 14] the particles deposit uniformly across the wetted area. Indeed, the deposit pattern could alter substantially depending upon the system used and some examples of the patterns include, a uniform deposition [15], an inner deposit [16], a stick-slip pattern resulting in multiple rings [17], fingering-like pattern [3], patterns displaying cracks [18], a crystalline pattern [19], a ring with an inner deposit [10] and patterns combining more than one of these features.

There have been numerous previous studies reporting the effect of particle size on the deposit patterns. For example, Deegan [20] showed that the evaporation of an aqueous polystyrene particles suspension on mica substrate produce ring with smaller inner rings for 0.1 $\mu$m particles while three zones, namely, ring, arches and radial lines with half arches form for 1 $\mu$m particles.

Chon et al. [21] studied evaporation of droplet containing different diameters of particles, namely, Au (2 nm), $Al_2O_3$ (11 nm and 47 nm), and CuO (30 nm), with 0.5 % concentration. They found that the droplet containing smaller particles deposit in a wider ring with a central deposition while a droplet containing larger particles corresponds to a distinct narrower ring. Perelaer et al. [22] investigated the deposition behavior of silica particles of different sizes (0.33 $\mu$m to 5 $\mu$m) and reported that the diameter of the dried deposit is slightly smaller than the initial wetted diameter and difference between the two diameters increases with an increase in particle diameter due to the wedge shape of the interface at the contact line. Bhardwaj et al. [8] studied evaporation of colloidal droplet containing 0.1 and 1 $\mu$m polystyrene particles at 1.0 % particles concentration. They reported torus-like ring profiles for smaller particles size, whereas monolayer ring was formed for larger particle size. Weon and Je [23] explored self-pinning characteristics of a decalin droplet suspended with PMMA particles and found that the colloidal droplet pins at the initial contact line irrespective of particles size. The capillary force experienced by the particles at the contact line retards the spreading of the droplet due to pinning. Subsequently [3], they showed that small PMMA colloids of $d = 0.1$ $\mu$m tend to form coffee-ring at contact line, whereas, large colloids of $d = 1.0$ $\mu$m tends to form an inner deposit for the decalin droplet and with an increase in particles concentration, the ring width, and diameter of the inner deposit increases. Yang et al. [24] recorded different deposits - multi-rings, radial spokes, spider web, foam, and islands - using droplets containing sulfate-modified polystyrene particles of different size ([20-200] nm) and concentrations over a very small range ([0.1−0.5] %v/v). They attributed these deposits to competition between the receding contact line velocity and the particle deposition rate at the contact line. Ryu et al. [25] observed rings for 0.1, and 1.0 $\mu$m particles and bumps for 5 and 10 $\mu$m particles. The absence of flocculation of the larger particles in the droplet hinders ring formation. Patil et al. [4] studied the deposition patterns of polystyrene particles of different sizes (0.1, 0.46, 1.1, and 3 $\mu$m) on a silicon wafer. They reported that an inner deposit and a thin ring with inner deposit form for smaller and larger particles, respectively. The formation of the deposits was explained by early depinning and self-pinning of the contact line in the two cases, respectively. Hence, a wide range of behavior has been identified as a result of changing particle size, but these studies consider only a single concentration or a very limited range of concentrations.

In the context of the effect of particles concentration on the ring-like deposits, Orejon et al. [26] reported that suspended $TiO_2$ nanoparticles in a water droplet induce stick-slip motion of the contact line and that with an increase in the TiO$_2$ concentration, the depinning time of the contact line increases on both hydrophilic and hydrophobic surface. Brutin [27] observed a ring without any inner deposits for a critical concentration of 1.15 vol% of 24 nm size polystyrene particles, and above that, a flower-like inner deposit was observed. Nguyen et al. [28] observed the formation of inner deposit with organic pigment nanoparticles during the evaporation of sessile water droplets on smooth hydrophobic surfaces. The radius of this inner deposit is smaller than the initial wetted radius and as it increases with an increase in the particle concentration. Ryu et al. [25] reported that the increase in the concentration of polymer (PEO) to the water colloidal solution of PMMA, suppresses the ring in case of smaller colloids, however, enhances in case of larger colloids. The aggregation of the small colloids with the addition of polymer prevents the outward radial flow of the particles, however, for large colloids, ring-induced hydrodynamics is more predominant than colloid-polymer interactions [25]. Lee et al. [29] observed uniform deposits of Al$_2$O$_3$ particles (0.1 $\mu$m) on a glass substrate at concentrations of larger than equal to 1 %. They reported that the ring forms on a less wettable surface and the ring width is larger for smaller particles. In a recent study, Patil et al. [10] measured ring profiles for 460 nm particles at 0.1 % and 1 % concentration and reported an increase in the ring width and height with the concentration. Sondej et al. [30] used white light interferometry to measure the ring profiles obtained after evaporation of a sessile water droplet containing sodium benzoate particles and reported an increase in ring height and reduction of drying rate with an increase in the particles concentration. Again, while there is a large number of studies on the effect of concentration, they focus on a single particle size.

Regarding the morphology of the ring, cracks on the surface of the ring have been reported in previous studies. For example, Pauchard et al. [31] observed a radial crack pattern at regular intervals on the ring of the dried deposit of aqueous silica solution (particles diameter of around 30 nm) with the addition of small quantity of salt. Zeid et al. [32] controlled the relative humidity to obtain a larger evaporation rate and reported radially ordered cracks on the ring of a dried blood droplet. Zhang et al. [33] reported radial crack patterns on the ring of PTFE deposits with an increase in crack length and crack spacing with an increase in PTFE concentration. They also reported that the surface wrinkling of the gel phase of the drying droplet to be one of the reasons for the initiation of cracks. Kim et al. [34] studied the crack formation mechanism on drying of a colloidal droplet of PMMA particles and observed that the crack initiation mechanism is favorable in small colloids (0.1 $\mu$m diameter) than that to the large colloids (1 $\mu$m diameter) due to air invasion inside the ring in the large colloids case. Dugyala et al. [35] reported on the effect of particles shape on the cracks formed in the ring and observed radial and concentric crack patterns for spherical and ellipsoidal particles, respectively. Lama et al. [18] studied crack formation in the ring using silica nanoparticles and polystyrene particles (0.1 $\mu$m diameter) of different concentration of [0.1-2.0] % wt and observed larger crack density and ordered cracks on the ring at a larger substrate temperature due to lesser ring height. The particle hydrophobicity can also influence the morphology, as reported by Shao et al. [36]. They showed that the hydrophobic particles result in a spoke-like deposit as compared to a typical ring-like deposit due to stronger capillary forces among them [36]. Here, too though, there has been little work on examining the occurrence of cracks in a single system over a wide range of concentrations and particle sizes, or quantitative studies of the rings formed.

In summary, while the influence of either particle size or concentration on the deposit formation has been studied, there has been no paper reporting on the coupled effect of these two key parameters, to the best of our knowledge, with exception of the work of Weon and Je [3]. However, in this paper, authors investigated the deposition patterns for a system (decalin/glass) that exhibits Marangoni effect while the focus of the present work is to consider a system (water/glass) without Marangoni effect. Note that the evaporating water droplets in ambient temperature do not exhibit Marangoni effect [37]. The second issue in this arena is that the quantitative measurements of ring profile or dimensions as a function of particle size and concentration have not been reported to the best of our knowledge, with an exception of the work of Sondej et al. [30]. In this study, quantitative measurements of the ring profiles were reported at different particles concentration [30]. The present work investigates the cross-sectional profile of the ring as a function of particle size and concentration. In the context of the ring morphology, it is not clear under which conditions the deposit would comprise of mono or multiple layers of particles and how cracks on the surface of the ring-like deposit would be influenced by particles size and concentration. It has been shown in the literature that the particle size influences the deposit pattern and shape significantly. For instance, larger colloids (20 $\mu$m polystyrene microspheres) deposit inside the ring as compared to smaller colloids (2 $\mu$m polystyrene microspheres) that deposit in the ring, explained by a larger capillary force on the larger colloids [38]. In addition, it is well-known that larger colloids exhibit strong capillary forces among them during drying [39]. However, the monolayer ring formed for large colloids below a critical particle concentration and the dynamics in its formation have not been explored in the literature. Therefore, the overall objective of the present paper is to quantitatively investigate the coupled effect of the particles size and concentration on the profile, dimensions, and morphology of the ring.

## 2. Experimental methods

Aqueous colloidal suspensions of 10 % v/v of uniformly monodispersed polystyrene latex beads of diameter $d$ = 0.1 μm (LB1), 1.1 μm (LB11), and 3.0 μm (LB30) were obtained from Sigma Aldrich Inc. The standard deviation of the particles diameter in the suspension is on the order of 5-15% of the mean diameter and the particles density is around 1005 kg/m$^3$, as per the manufacturer's data sheet. Solutions over a wide range of concentrations, $c$ = 0.001, 0.01, 0.1, and 1.0 % v/v, were prepared by diluting with deionized water. After the dilution, we performed sonication for about 30 minutes to ensure uniform suspension of the particles in the solution. The droplets of the colloidal suspensions were generated using a micropipette (Prime, Biosystem Diagnostics Inc.) of the volume of 1.1 ± 0.2 μL. The droplets were gently deposited on the substrate. Glass slide (Sigma Aldrich, S8902) with dimensions of 75 x 25 x 1 mm$^3$ served as the substrate in all experiments. The glass slide was washed with isopropanol and was allowed to completely dry in the ambient conditions for few minutes before the droplet was deposited on it. A fresh slide was used to repeat or to perform a new experiment. The droplet wetted radius ($R$) and height ($H$) were measured using images obtained from a side view. Since the wetted diameter of the deposited droplet was below the capillary length of water ($\approx$ 2.7 mm), the droplet displays a spherical cap and the initial static contact angle is given by, $\theta_c = 2\tan^{-1}(H/R)$ [40]. The measured values of the contact angle and wetted radius based on three experimental runs are listed for all experiments are provided in the supporting information (Tables S1 and S2). The uncertainties in these measurements are around ±1° and ± 0.1 mm, respectively. While the measured initial static contact angle does not show significant variation with particle concentrations, there is a slight increase in the contact angle as particle size increases (around 22% increase for $d$ = 3 μm as compared to $d$ = 0.1 μm). This could be due to a different amount of surfactant in the suspensions of different particle sizes, used by the manufacturer for stabilization of the suspensions. Note that the increase in the contact angle for $d$ = 3 μm corresponds to a decrease in the wetted radius (Table S3).

The dried patterns of the colloidal particles were visualized from the top by an optical microscope (Olympus Corp. Inc., BX53F, with a magnification of 10X to 40X). During the evaporation, particle motion was visualized by a high-speed camera (IDT Inc, Motion- Pro Y-3 classic) mounted on the optical microscope. Field emission gun scanning electron microscope (JSM-7600F, Jeol Inc) was employed to record a high-resolution view of the ring morphology. The ring profiles were quantitatively measured by a 3D optical profilometer (Zeta-20, Zeta Instruments Inc., optical resolution ~ 0.1 $\mu$m). The ring profiles were measured at four azimuthal locations (left, top, right, and bottom side) on the ring, as shown in Figure 1(b). The measured profiles for a representative case ($d = 1.1$ μm, and $c = 0.1$ %) are plotted in Figure 1(c). An average profile was obtained using the profiles at these four locations, as shown by a thick black line in Figure 1(c). The maximum estimated uncertainty in the measured height due to the averaging and run-to-run variation is around ±10%. All experiments were performed three times to ensure repeatability. The ambient temperature and relative humidity were 27 ± 2 °C and 35 ± 5%, respectively.

## 3. Results and discussion

We present results for the evaporation of 1.1 ± 0.2 $\mu$L water droplets containing polystyrene particles on hydrophilic glass substrate for three cases of particle diameter ($d = 0.1, 1.1, 3.0$ $\mu$m) and four cases of particle concentrations ($c = 0.001, 0.01, 0.1$ and $1.0$ %v/v). These droplets evaporate in constant contact radius (CCR) mode and the droplet evaporation characteristics have been quantified in our previous work [10].

### 3.1. Deposit patterns and morphology of the ring

Figure 2 shows the top view of dried deposits obtained after the evaporation of the droplet for different cases of particle size and concentration. Each row and column in Figure 2 represent a constant particle size and constant particle concentration, respectively. The deposit is predominantly a ring in all cases and its formation is explained as follows. The contact line remains pinned during the evaporation for the aqueous droplet considered here, as reported in our previous work [10]. The ring forms due to advection of the particles by the evaporation-driven outward flow [1]. We confirm the flow direction by recording the particle motion using a microscope (see videos SI.1, SI.2, and SI.3 provided in the supporting information). As reported by Hu and Larson [7], the Marangoni flow is absent during evaporation of a water droplet at ambient temperature, consistent with our observations here. Note that the substrate thickness is large enough not to induce a thermal gradient on the liquid-gas interface due to the latent heat of evaporation [41]. At larger particles concentration and larger particle diameter ($c = 1.0$ % for $d = 1.1$ $\mu$m and 3 $\mu$m), alongside the formation of the ring, particles are also deposited in the inner region of the droplet. The settling velocity of the particles can be estimated using Stokes law as follows,

$$U_s = \frac{d^2(\rho_P - \rho)g}{18\mu} \tag{1}$$

where $d$, $\rho_p$, $\rho$, $g$ and $\mu$ are particles diameter, particle density, water density, gravitational acceleration, and viscosity, respectively. The evaporation-induced advection velocity scales as follows [11],

$$U_e = \frac{j_{max}}{\rho} \tag{2}$$

where $j_{max}$ is the maximum evaporation mass flux [kg m$^{-2}$ s$^{-1}$] near the pinned contact line. The estimated values of $U_e/U_s$ for $d = 0.1$, 1.1, and 3.0 $\mu$m using eqs. 1-2 are $2.5\times10^4$, $2.2\times10^2$, and 24, respectively. $j_{max}$ is estimated using the analytical expression given by Hu and Larson [42]. This implies that the evaporation-induced advection overwhelms gravitational sedimentation in all cases of particle diameter considered here. Therefore, the main mechanism of the particles deposition in the ring in the present work is the same as reported in classical "coffee-ring" deposition [1]. The ring width increases with the increase in the concentration for a constant particle size as observed qualitatively in Figure 2. We also note that the particles deposit in the inner region of the wetted area at large particles concentration ($c = 1\%$, last column in Figure 2).

Figure 3 shows a zoomed-in view of the ring, exhibiting morphology, for all cases. The contact line is on the left of each frame. Note that few out-of-focus marks seen in some images (specifically for $d = 1$ $\mu$m) are experimental artifacts. In Figure 3 (second row) for $d = 1.1$ $\mu$m, a discontinuous and continuous monolayer ring are observed at $c = 0.001$, and 0.01%, respectively (the ring height is quantified in section 3.2). Similarly, at $d = 3.0$ $\mu$m and for $c = 0.001$ and 0.01, discontinuous monolayer ring forms while for $c = 0.1\%$, continuous monolayer ring forms. As $c$ increases, the particles stack up in the ring at $c = 0.1$, and 1.0 % for $d = 1.1$ $\mu$m and at $c = 1.0\%$ for $d = 3.0$ $\mu$m and the rings with multiple layers of the particles form. In other words, with an increase of particle concentration more particles are deposited in the ring, resulting in it being both thicker and more densely packed, this is observed qualitatively in Figure 3. The formation of the monolayer ring can be attributed to the increase in the inter-particle capillary forces at larger particle size, as these forces scale with $d^2$ [39]. The deposits of $d = 1.1$ $\mu$m (Figure 3, $c = 0.01$ to 1%) and $d = 3$ $\mu$m (Figure 3, $c = 0.1$ to 1% and Figure 4(b)) show ordered crystal-like morphology near the contact line while disordered random aggregates form closer to inner boundary of the ring and in the inner region. As explained by Marin et al. [43], these different morphologies exist due to a rather slow deposition of the particles in the initial stage of the evaporation, due to which particles get enough time to order themselves by Brownian motion into a crystal-like structure.

For more complex stacked structures which can be deposited, under certain conditions, cracking occurs during ring formation for $d = 0.1$ $\mu$m at $c = 0.1$ % and $c = 1.0$ % (last two frames of the first row of Figure 3). At $c = 0.1$ %, disordered and dendritic-like cracks are obtained, while for $c = 1.0$ % the cracks are ordered along the radial direction and aligned along the direction of the drying front. A SEM image of the cracks at $d = 0.1$ $\mu$m, $c = 0.1$ % (Figure 4(a)) shows a dense packing of the particles at the two interfaces of the crack. As pointed to in Refs. [18,35,44,45], a growing ring exhibits a gel-like behavior with a dense packing of the particles. As the liquid evaporates in the final drying stage, the ring tries to shrink, and the pinned contact line obstructs such shrinkage. This results in stresses in the deposit and consequently, it induces cracks in the ring. The appearance of the cracks only in the last stages of the evaporation is confirmed by the video (SI.1) showing particles deposition in the ring for $d = 0.1$ $\mu$m, $c = 0.1$ %. The number of cracks per unit length for $d = 0.1$ $\mu$m, $c = 0.1$, and 1.0 % are estimated from the zoomed-in microscopy images shown in Figure 3 and are 0.08 and 0.06 $\mu$m$^{-1}$, respectively (see Table S3 in supporting information). The spacing between two consecutive cracks increases with an increase in the particles concentration, consistent with previously reported data by Dugyala et al. [35] for ellipsoidal particles. The cracks are not observed for larger colloidal particles, as confirmed by SEM image for $d = 1.1$ $\mu$m in Figure 4(b). This can be explained by the fact that the critical cracking liquid film thickness ($h_{crit}$) is larger for larger colloidal (hard) spheres ($h_{crit} \sim d^{3/2}$) [46]. This result is also consistent with results of Kim et al. [34], in which 1 $\mu$m PMMA colloids suspended in decalin do not show cracks as compared to 0.1 $\mu$m particles.

We compare dynamics of the particle deposition process for different cases of particle size and plot time-sequence images recorded by microscopy showing time-varying particles deposition in the ring for particle sizes of $d = 0.1$, 1.1 and 3.0 $\mu$m in Figure 5(a), (b) and (c), respectively, at $c = 0.1$ % (see also associated movies for these cases, SI.1, SI.2, and SI.3). The focal plane is on the substrate surface in all cases and out-of-focus particles are visible in case of $d = 1.1$ and 3.0 $\mu$m in Figure 5(b) and (c), respectively. The contact line is pinned during the evaporation and particles advect towards the ring in all cases. The ring front shows a significant growth for 0.1 $\mu$m and 1.1 $\mu$m particles in Figure 5(a) and 5(b), respectively and the particles stack up in multiple layers in these two cases (also confirmed in Figure 4). The cracks are visible on the ring surface at $t_0 + 150$ s for 0.1 $\mu$m particles in Figure 5(a) and Figure 4(a). In Figure 5(c), 3.0 $\mu$m particles form a monolayer of particles in the ring instead of stacking up in the multiple layers. The video of recorded motion of the particles (SI.3) shows that the particles are dropping into focus before getting to the contact line and are being delivered into a monolayer ring. During initial stages of the evaporation ($t < t_0$), the particles are blocked in a wedge-like region of the contact line and deposits few micrometers away from the contact line (Figure 5(c)), consistent with findings reported by Patil et al. for 3.0 $\mu$m particles [4]. At later times ($t > t_0 + 40$ s), other incoming particles to the contact line adhere to the ones, which are already present near the contact line due to large immersion capillary force among the particles, that scales as $d^2$ [39]. In the last stages of the evaporation, some particles also deposit in the inner region as the contact line recedes and the remaining liquid film dries out.

### 3.2. Measurement of ring profiles

Figure 6 compares the measured ring profiles for different particle concentration ($c = 0.001$, 0.01, 0.1 and 1.0 %v/v), keeping particle size constant. Note that the origin is located at the outer periphery of the ring. In Figure 6(a), at $d = 0.1$ $\mu$m, the ring profile resembles a partial torus-like shape and such a profile is attributed to a build-up of particles at the contact line due to the outward flow in the initial stages of the evaporation [1] and depinning of the contact line from growing deposit in last stages of the evaporation. The ring height ($h$) and width ($w$) increase with the particle concentration ($c$). Since the maximum ring height ($h$) is larger than the particle size in all cases of concentration, the particles are clearly stacked in multiple layers in the ring. The increase in the ring width with concentration at $d = 0.1$ $\mu$m is qualitatively confirmed in the first row of Figure 3. In Figure 6(b), at $d = 1.1$ $\mu$m, the maximum ring height is almost equal to particle size ($h \approx d = 1.1$ $\mu$m) for $c = 0.001$, and 0.01 %. However, for $c = 0.1$, and 1.0 %, the ring height is larger than the particle size ($h > 1.1$ $\mu$m). Therefore, the particles form a monolayer (a single layer of the particles) in the ring at $c = 0.001$, and 0.01 % for $d = 1.1$ μm. The monolayer is discontinuous and continuous at $c = 0.001$, and 0.01 %, respectively, as confirmed from optical microscopy images in Figure 3 (first two frames in the second row). At $c = 0.1$ and 1.0 %, the ring width increases with an increase in particle concentration. In Figure 6(c), at $d = 3.0$ μm, the maximum ring height ($h$) is same as the particle diameter for $c = 0.001$, 0.01, and 0.1 %, corresponding to a monolayer ($h \approx d$). At $c = 1.0$ %, the ring height and width increase with an increase in the particle concentration, qualitatively shown in the third row of Figure 3. The monolayers are discontinuous for $c = 0.001$, 0.01 % while the monolayer is continuous for $c = 0.1$ %. The mechanism of the formation of the monolayer in Figure 6(b) and (c) at low concentration was explained earlier in section 3.1. However, for larger concentrations at larger particle sizes, there is a sufficient number of particles available in the droplet which stack up as multiple layers near the contact line.

### 3.3. Regime map

We plot a regime map to classify monolayer and multiple layers ring formation on particle concentration - particle size plane in Figure 7. The discontinuous or continuous monolayer ring occur for larger particle size combined with low concentration, while the multiple layers ring form at smaller particle size and larger concentration. A dashed line demarcates qualitatively the three regimes and the cracks form in two cases of multiple layers, as shown in Figure 7. Our measurements show the existence of a critical concentration at a constant particle diameter or a critical particle diameter at a constant particle concentration, for the formation of the monolayer ring.

### 3.4. Scaling of ring dimensions with particles concentration

We compare scaling of the ring dimensions with predictions of a model, proposed by Popov [47] and based on the conservation of the mass of droplet liquid and particles during the evaporation. The model treats the particles as continua and ignores immersion capillary forces among the particles during the formation of the monolayer ring. Therefore, we do not compare data of $d = 1.1$ μm and $d = 3.0$ μm against the model since monolayer rings form in this case at low concentrations. In this model, the non-dimensional ring width ($W$) and height ($H$) are expressed, respectively, as follows,

$$W = w/R = 0.6\sqrt{(c/p)} \qquad (3)$$

$$H = h/R = 0.3\theta_c \sqrt{(c/p)} \quad (4)$$

where $w$, $h$, $c$, $p$, $\theta_c$, $R$ are ring width, ring height, particle concentration, particle packing fraction, static contact angle, and initial droplet wetted radius, respectively.

Figure 8(a) and (b) shows the qualitative comparison between the present measurements for $d = 0.1$ $\mu$m and model predictions for $W$ and $H$, respectively, as a function of particle concentration ($c$) on a log-log scale. The measurements of $d = 1.1$ $\mu$m and $d = 3.0$ $\mu$m including those result in the monolayer are also plotted in Figure 8. The model predicts the same value of $W$ for all particle sizes, while $H$ varies slightly with size due to the dependence of the $H$ on $\theta_c$ (eq. 4), due to the assumption of a wedge-like contact line region in the model (Figure 9(a)). The measured width and height of the ring scales non-linearly with particle concentration i.e. $W \sim c^m$ and $H \sim c^n$. The values of $m$ and $n$ obtained using least squares curve fitting method for $d = 0.1$ $\mu$m are 0.55 and 0.41, respectively, both values are closer to 0.5, as predicted by the model.

Interestingly, the model underpredicts and overpredicts the width (Figure 8(a)) and height (Figure 8(b)) as compared to the measurements, respectively. This is attributed to the fact that the model presents the shape of the ring profile as a wedge-like region near the contact line (Figure 9(a)) and it does not account for the interaction of the shrinking free surface with the growing deposit in the last stage of the evaporation [8], which results in a typical partial torus-like profile of the ring. To verify this hypothesis, we compare the mass of the particles in the ring obtained in the measurements ($M_{\text{ring, exp}}$) and predicted by the model ($M_{\text{ring, model}}$) at different particles size and concentrations for multiple layer ring cases. The mass of the ring ($M_{\text{ring, exp}}$) is estimated by the following expression,

$$M_{\text{ring, exp}} = \int_{R-w}^{R} \rho_p p f(r) 2\pi r dr \quad (5)$$

where $\rho_p$, $p$, $f(r)$, $R$, $w$ are particles density, particles packing fraction (or deposit porosity), ring profile, wetted radius, and ring width, respectively. The ring profile ($f(r)$) is obtained by fitting a second-order polynomial curve using the least-square fitting method to the measured ring profile. The value of $p$ is taken as 0.656 from Ref. [47]. We estimated around ±20% uncertainty in calculated $M_{\text{ring, exp}}$, based on the uncertainties in the measured ring height and droplet volume. $M_{\text{ring, model}}$ was calculated by approximating the ring to a wedge-like profile and by integrating it over the wetted radius ($R$). We obtain the following expression after the integration,

$$M_{\text{ring, model}} = \rho_p p \pi R w h \tag{6}$$

Figure 9(b) shows the comparison between $M_{\text{ring, exp}}$, and $M_{\text{ring, model}}$ at different particles size and concentrations for cases of multiple layer ring. Since monolayer ring cases cannot be predicted by a continuum based model, we did not plot these cases in Figure 9(b). The measured ring mass is very close to the mass predicted by the model and the scaling of the measured ring mass with particles concentration is captured by the model. Thus, the difference in the predicted and measured ring dimensions in Figure 8(a) and (b) is due to the spreading of the ring at the expense of its height in the last stage of the drying, explained by the interaction of the growing deposit with shrinking free surface (Figure 9(a)). The interaction is not captured by the model proposed by Popov [47].

We further verify the hypothesis by comparing dynamics of the deposition of the particles for the formation of the monolayer for $d = 3$ $\mu$m, with predictions of a model, proposed by Deegan et al. [48]. In this model, the number of particles migrating towards the contact line at a given time, $t$, follows a power law given by, $N \sim t^{2/1+\lambda}$ [48], where $\lambda$ is a function of the initial static contact angle ($\theta_c$), $\lambda = (\pi-2\theta_c)/(2\pi-2\theta_c)$. We count the number of particles migrating towards the contact line for case $d = 3.0$ $\mu$m and $c = 0.1$ % (visualization of the particle motion of this case is plotted in Figure 5). Since the counting of smaller colloids ($d = 0.1$ and $1.1$ $\mu$m) was not possible in recorded particle motion (Figure 5(a, b)), we did not plot this data for the smaller colloids. Figure 9(c) shows that the present measurements agree well with the $N$-$t$ relationship predicted by the model, where $\lambda = 0.35$, at $t \leq 80$ s. The model does not account for the interaction of the growing deposit with shrinking surface and the difference between the model prediction and measurement start to increase at $t > 80$ s. As explained earlier, the ring spreads at the expense of its height during this interaction. The contact line recedes at $t \approx 120$ s, resulting in the formation of a monolayer of particles in the ring.

## 4. Conclusions

We have studied ring-like deposits obtained after the evaporation of a sessile water droplet containing polystyrene colloidal particles on a glass substrate. The coupled effect of the particle size ($d$) and particles concentration ($c$) on ring dimensions and morphology have been investigated. The range of $d$ and $c$ in the experiments are [0.1, 3] $\mu$m and [0.001, 1] %v/v, respectively. The dried patterns were visualized under an optical microscope and the ring profiles were measured by an optical profilometer. The measured ring profiles resemble a partial torus-like shape for all cases of $d$ and $c$. We have visualized cracks on the surface of the ring for $d = 0.1$ $\mu$m and $c = 0.1$, and 1.0 % and have explained their formation briefly. Three types of deposits are classified on particles concentration - particle size plane, namely, discontinuous monolayer ring, continuous monolayer ring, and multiple layers ring. In the case of multiple layers, the ring width and height increases with an increase in particles concentration and relative mass of the particles accumulated in the ring is the largest at the lowest particle size. The monolayer ring forms due to larger immersion capillary forces among the particles at larger particle size and measured particle motion near the contact line qualitatively confirm the interaction among the particles. The present measurements show that a critical particle concentration may exist at a given particle size to achieve the formation of the monolayer ring. The qualitative measured variation of the ring dimensions with particles concentration is consistent with the predictions of an existing theoretical model at $d = 0.1$ $\mu$m. The measured dimensions of the ring scale with particles concentration by a power law and in general, the scaling agrees with the predictions of the model. We compare time-varying ring dimensions with the model and conclude that a growing ring spreads in the last stage of drying at the expense of its height due to its interaction with shrinking free surface.

Overall, the present study provides fundamental insights into self-assembly of colloidal particles and dependence of the ring profile and dimensions on particle size and particles concentration for microliter aqueous droplets at low particles concentration. The present results are potentially useful to design technical applications such as manufacturing of bioassays and biosensors. However, note that the droplets employed here are bigger (microliter) as compared to smaller droplets (picoliter) used in inkjet printing. In addition, the maximum particle concentration considered in the present work is 1% while larger particles concentration is typically used in surface coating applications (for instance, see discussion by Sondej et al. [30]).

## 5. Supporting information

Videos (AVI) of visualization of particle motion near contact line of the evaporating droplet for three cases of particles diameter, $d = 0.1$ μm (SI.1), 1.1 μm (SI.2) and 3.0 μm (SI.3), for 0.1% particles concentration. Measurements of contact angle, wetted radius and crack spacing (PDF).

## 6. Acknowledgments

R.B. gratefully acknowledges financial support by a grant (EMR/2016/006326) from Science and Engineering Research Board (SERB), Department of Science and Technology (DST), New Delhi,

## 7. References

[1] R.D. Deegan, O. Bakajin, T.F. Dupont, G. Huber, S.R. Nagel, T.A. Witten, Capillary flow as the cause of ring stains from dried liquid drops, Nature. 389 (1997) 827–829. doi:10.1038/39827.

[2] R.G. Larson, In Retrospect: Twenty years of drying droplets, Nature. 550 (2017) 466–467. doi:10.1038/550466a.

[3] B.M. Weon, J.H. Je, Fingering inside the coffee ring, Phys. Rev. E - Stat. Nonlinear, Soft Matter Phys. 87 (2013) 1–6. doi:10.1103/PhysRevE.87.013003.

[4] N.D. Patil, R. Bhardwaj, A. Sharma, Self-sorting of Bi-dispersed Colloidal Particles near Contact Line of an Evaporating Sessile Droplet, Langmuir. (2018). doi:10.1021/acs.langmuir.8b00427.

[5] A.F. Routh, Drying of thin colloidal films, Reports Prog. Phys. 046603 (2013). doi:10.1088/0034-4885/76/4/046603.

[6] M. Parsa, S. Harmand, K. Sefiane, Mechanisms of pattern formation from dried sessile drops, Adv. Colloid Interface Sci. (2018). doi:10.1016/j.cis.2018.03.007.

[7] H. Hu, R.G. Larson, Marangoni effect reverses coffee-ring depositions, J. Phys. Chem. B. 110 (2006) 7090–7094. doi:10.1021/jp0609232.

[8] R. Bhardwaj, X. Fang, D. Attinger, Pattern formation during the evaporation of a colloidal nanoliter drop: A numerical and experimental study, New J. Phys. 11 (2009). doi:10.1088/1367-2630/11/7/075020.

[9] Y. Li, C. Lv, Z. Li, D. Quéré, Q. Zheng, From coffee rings to coffee eyes, Soft Matter. 11 (2015) 4669–4673. doi:10.1039/C5SM00654F.

[10] N.D. Patil, P.G. Bange, R. Bhardwaj, A. Sharma, Effects of Substrate Heating and Wettability on Evaporation Dynamics and Deposition Patterns for a Sessile Water Droplet Containing Colloidal Particles, Langmuir. 32 (2016) 11958–11972. doi:10.1021/acs.langmuir.6b02769.

[11] R. Bhardwaj, X. Fang, P. Somasundaran, D. Attinger, Self-assembly of colloidal particles from evaporating droplets: Role of DLVO interactions and proposition of a phase diagram, Langmuir. 26 (2010) 7833–7842. doi:10.1021/la9047227.

[12] P.J. Yunker, T. Still, M.A. Lohr, A.G. Yodh, Suppression of the coffee-ring effect by shape-dependent capillary interactions, Nature. 476 (2011) 308–311. doi:10.1038/nature10344.

[13] H.B. Eral, D.M. Augustine, M.H.G. Duits, F. Mugele, Suppressing the coffee stain effect: How to control colloidal self-assembly in evaporating drops using electrowetting, Soft Matter. 7 (2011) 4954–4958. doi:10.1039/c1sm05183k.

[14] D. Orejon, K. Sefiane, M.E.R. Shanahan, Evaporation of nanofluid droplets with applied DC potential, J. Colloid Interface Sci. 407 (2013) 29–38. doi:10.1016/j.jcis.2013.05.079.

[15] T.P. Bigioni, X.M. Lin, T.T. Nguyen, E.I. Corwin, T.A. Witten, H.M. Jaeger, Kinetically driven self assembly of highly ordered nanoparticle monolayers, Nat. Mater. 5 (2006) 265–270. doi:10.1038/nmat1611.

[16] J. Wu, J. Xia, W. Lei, B.P. Wang, Generation of the smallest coffee-ring structures by solute crystallization reaction on a hydrophobic surface, RSC Adv. 3 (2013) 5328–5331. doi:10.1039/c3ra40465j.

[17] S. Maheshwari, L. Zhang, Y. Zhu, H.C. Chang, Coupling between precipitation and contact-line dynamics: Multiring stains and stick-slip motion, Phys. Rev. Lett. 100 (2008) 1–4. doi:10.1103/PhysRevLett.100.044503.

[18] H. Lama, M.G. Basavaraj, D.K. Satapathy, Tailoring crack morphology in coffee-ring deposits via substrate heating, Soft Matter. (2017). doi:10.1039/C7SM00567A.

[19] D. Kaya, V.A. Belyi, M. Muthukumar, Pattern formation in drying droplets of polyelectrolyte and salt, J. Chem. Phys. 133 (2010). doi:10.1063/1.3493687.

[20] R. Deegan, Pattern formation in drying drops, Phys. Rev. E. 61 (2000) 475–485. doi:10.1103/PhysRevE.61.475.

[21] C.H. Chon, S. Paik, J.B. Tipton, K.D. Kihm, Effect of nanoparticle sizes and number densities on the evaporation and dryout characteristics for strongly pinned nanofluid droplets, Langmuir. 23 (2007) 2953–2960. doi:10.1021/la061661y.

[22] J. Perelaer, P.J. Smith, C.E. Hendriks, A.M.J. van den Berg, U.S. Schubert, The preferential deposition of silica micro-particles at the boundary of inkjet printed droplets, Soft Matter. 4 (2008) 1072. doi:10.1039/b715076h.

[23] B.M. Weon, J.H. Je, Self-pinning by colloids confined at a contact line, Phys. Rev. Lett. 110 (2013) 1–5. doi:10.1103/PhysRevLett.110.028303.

[24] X. Yang, C.Y. Li, Y. Sun, From multi-ring to spider web and radial spoke: competition between the receding contact line and particle deposition in a drying colloidal drop, Soft Matter. 10 (2014) 4458–4463. doi:10.1039/C4SM00497C.

[25] S.A. Ryu, J.Y. Kim, S.Y. Kim, B.M. Weon, Drying-mediated patterns in colloid-polymer suspensions, Sci. Rep. 7 (2017) 1–7. doi:10.1038/s41598-017-00932-z.

[26] D. Orejon, K. Sefiane, M.E.R. Shanahan, Stick-slip of evaporating droplets: Substrate hydrophobicity and nanoparticle concentration, Langmuir. 27 (2011) 12834–12843. doi:10.1021/la2026736.

[27] D. Brutin, Influence of relative humidity and nano-particle concentration on pattern formation and evaporation rate of pinned drying drops of nanofluids, Colloids Surfaces A Physicochem. Eng. Asp. 429 (2013) 112–120. doi:10.1016/j.colsurfa.2013.03.012.

[28] T.A.H. Nguyen, A. V. Nguyen, M.A. Hampton, Z.P. Xu, L. Huang, V. Rudolph, Theoretical and experimental analysis of droplet evaporation on solid surfaces, Chem. Eng. Sci. 69 (2012) 522–529. doi:10.1016/j.ces.2011.11.009.

[29] H.H. Lee, S.C. Fu, C.Y. Tso, C.Y.H. Chao, Study of residue patterns of aqueous nanofluid droplets with different particle sizes and concentrations on different substrates, Int. J. Heat Mass Transf. 105 (2017) 230–236. doi:10.1016/j.ijheatmasstransfer.2016.09.093.

[30] F. Sondej, M. Peglow, A. Bück, E. Tsotsas, Experimental investigation of the morphology of salt deposits from drying sessile droplets by white-light interferometry, AIChE J. 64 (2018) 2002–2016. doi:10.1002/aic.16085.

[31] L. Pauchard, F. Parisse, C. Allain, Influence of salt content on crack patterns formed through colloidal suspension desiccation, Phys. Rev. E - Stat. Physics, Plasmas, Fluids, Relat. Interdiscip. Top. 59 (1999) 3737–3740. doi:10.1103/PhysRevE.59.3737.

[32] W.B. Zeid, J. Vicente, D. Brutin, Influence of evaporation rate on cracks' formation of a drying drop of whole blood, Colloids Surfaces A Physicochem. Eng. Asp. 432 (2013) 139–146. doi:10.1016/j.colsurfa.2013.04.044.

[33] Y. Zhang, Y. Qian, Z. Liu, Z. Li, D. Zang, Surface wrinkling and cracking dynamics in the drying of colloidal droplets, Eur. Phys. J. E. 37 (2014). doi:10.1140/epje/i2014-14084-3.

[34] J.Y. Kim, K. Cho, S. Ryu, S.Y. Kim, B.M. Weon, Crack formation and prevention in

colloidal drops, Sci. Rep. 5:13166 (2015) 1–9. doi:10.1038/srep13166.

[35] V.R. Dugyala, H. Lama, D.K. Satapathy, M.G. Basavaraj, Role of particle shape anisotropy on crack formation in drying of colloidal suspension, Sci. Rep. 6:30708 (2016) 1–7. doi:10.1038/srep30708.

[36] F.F. Shao, A. Neild, T.W. Ng, Hydrophobicity effect in the self assembly of particles in an evaporating droplet, J. Appl. Phys. 108 (2010) 1–8. doi:10.1063/1.3455845.

[37] H. Hu, R.G. Larson, Analysis of the effects of marangoni stresses on the microflow in an evaporating sessile droplet, Langmuir. 21 (2005) 3972–3980. doi:10.1021/la0475270.

[38] B.M. Weon, J.H. Je, Capillary force repels coffee-ring effect, Phys. Rev. E - Stat. Nonlinear, Soft Matter Phys. 82 (2010) 1–4. doi:10.1103/PhysRevE.82.015305.

[39] P.A. Kralchevsky, K. Nagayama, Capillary Forces between Colloidal Particles, Langmuir. 10 (1994) 23–36. doi:10.1021/la00013a004.

[40] R. Bhardwaj, J.P. Longtin, D. Attinger, Interfacial temperature measurements, high-speed visualization and finite-element simulations of droplet impact and evaporation on a solid surface, Int. J. Heat Mass Transf. 53 (2010) 3733–3744. doi:10.1016/j.ijheatmasstransfer.2010.04.024.

[41] M. Kumar, R. Bhardwaj, A combined computational and experimental investigation on evaporation of a sessile water droplet on a heated hydrophilic substrate, Int. J. Heat Mass Transf. 122 (2018) 1223–1238. doi:10.1016/j.ijheatmasstransfer.2018.02.065.

[42] H. Hu, R.G. Larson, Analysis of the microfluid flow in an evaporating sessile droplet, Langmuir. 21 (2005) 3963–3971. doi:10.1021/la047528s.

[43] Á.G. Marín, H. Gelderblom, D. Lohse, J.H. Snoeijer, Order-to-disorder transition in ring-shaped colloidal stains, Phys. Rev. Lett. 107 (2011) 1–4. doi:10.1103/PhysRevLett.107.085502.

[44] T. Okuzono, M. Kobayashi, M. Doi, Final shape of a drying thin film, Phys. Rev. E - Stat. Nonlinear, Soft Matter Phys. 80 (2009) 1–11. doi:10.1103/PhysRevE.80.021603.

[45] B. Sobac, D. Brutin, Desiccation of a sessile drop of blood : Cracks , folds formation and delamination, Colloids Surfaces A Physicochem. Eng. Asp. 448 (2014) 34–44. doi:10.1016/j.colsurfa.2014.01.076.

[46] K.B. Singh, M.S. Tirumkudulu, Cracking in Drying Colloidal Films, Phys. Rev. Lett. 218302 (2007) 1–4. doi:10.1103/PhysRevLett.98.218302.

[47] Y.O. Popov, Evaporative deposition patterns: Spatial dimensions of the deposit, Phys. Rev. E - Stat. Nonlinear, Soft Matter Phys. 71 (2005) 1–17. doi:10.1103/PhysRevE.71.036313.

[48] R.D. Deegan, O. Bakajin, T.F. Dupont, G. Huber, S.R. Nagel, T.A. Witten, Contact line deposits in an evaporating drop, Phys. Rev. E - Stat. Physics, Plasmas, Fluids, Relat. Interdiscip. Top. 62 (2000) 756–765. doi:10.1103/PhysRevE.62.756.

## 8. Figures

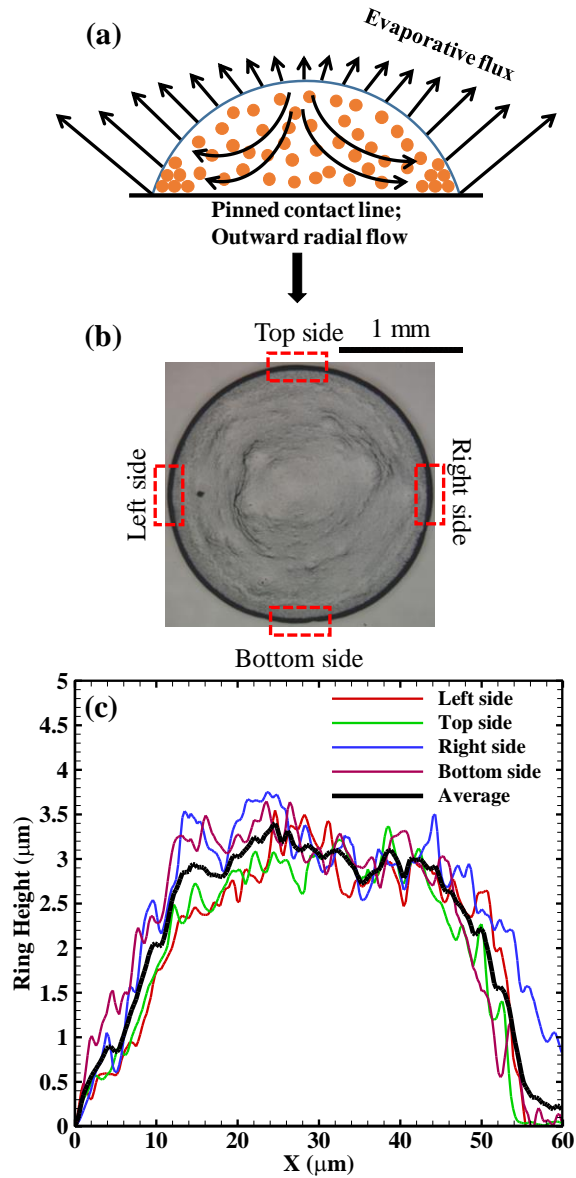

Figure 1. (a) Schematic representation of the evaporation-induced outward radial fluid flow and ring-like deposit on a hydrophilic substrate with a pinned contact line of the sessile droplet. (b) A typical ring-like deposit in the present measurement. The profile of the ring is measured at four locations (shown as a red dashed square) to get an averaged profile. (c) Measured ring profiles obtained after evaporation of 1.1 $\mu$L droplet containing 1.1 $\mu$m polystyrene particles with 0.1 %v/v concentration on a glass surface with an averaged profile (thicker black line).

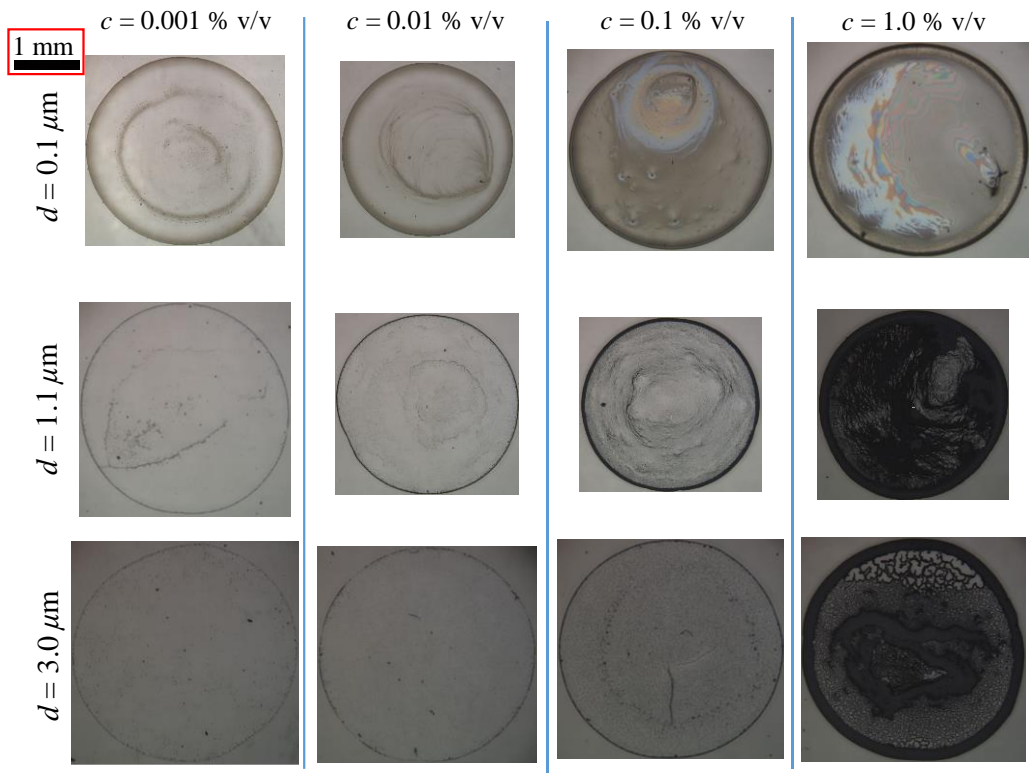

Figure 2. Ring-like deposit patterns recorded using optical microscopy and obtained after the evaporation of 1.1 µL water droplets containing polystyrene particles on a hydrophilic glass substrate. Particle size is kept constant in three rows ($d$ = 0.1, 1.1, 3.0 μm) and particle concentration is kept constant in four columns ($c$ = 0.001, 0.01, 0.1 and 1.0 %). The scale is shown on top left of the figure.

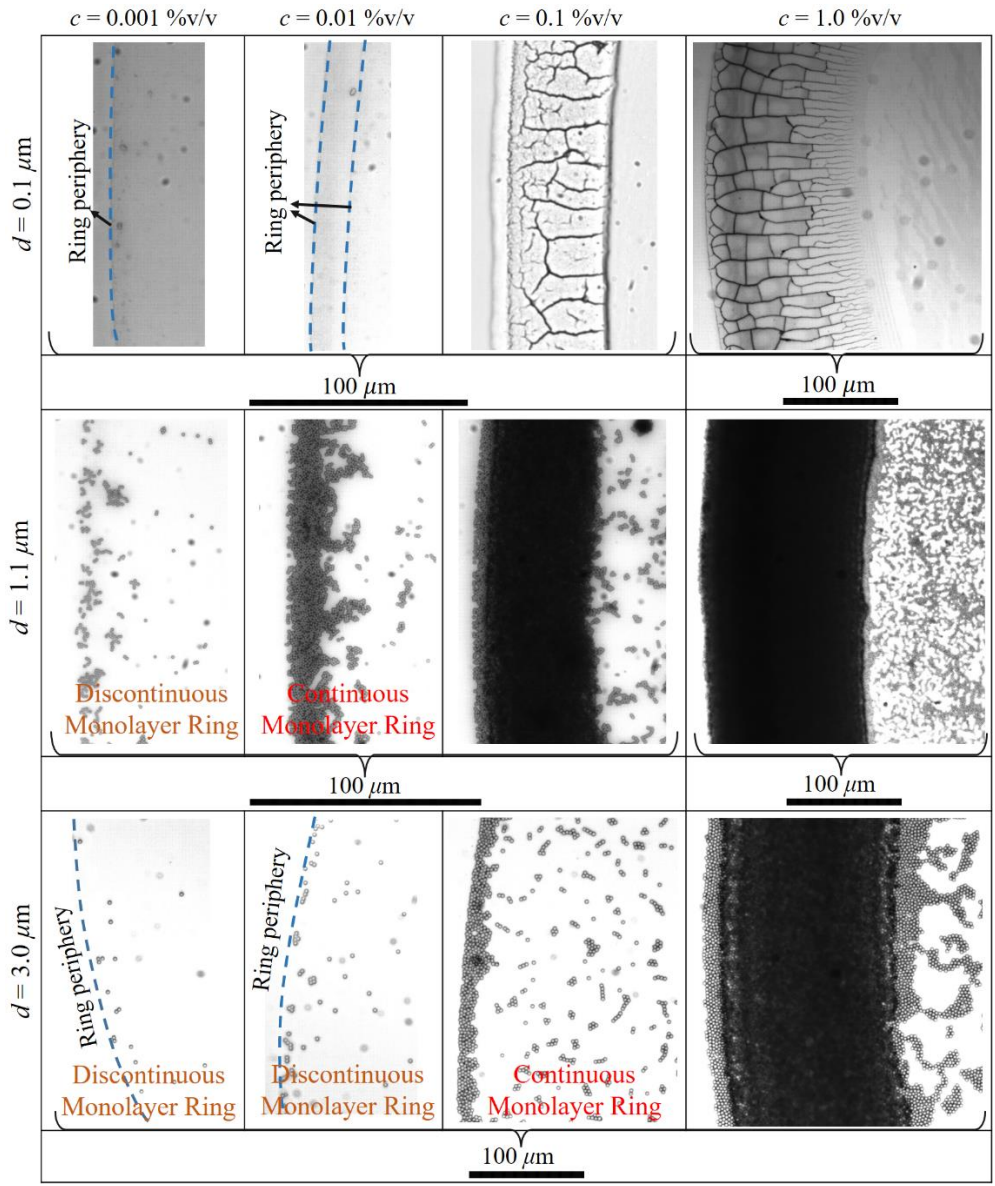

Figure 3. Morphologies of the rings for the cases plotted in Figure 2 are shown by the zoomed-in view of the ring for the respective case. Scale bars are given below the respective images.

**(a)** $d = 0.1$ μm, $c = 0.1$ % v/v

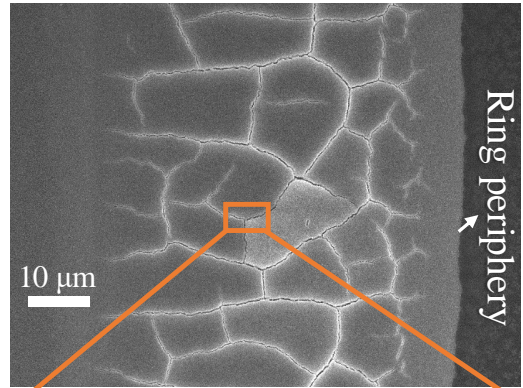

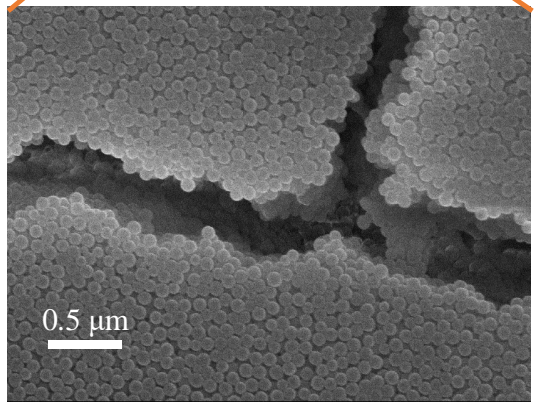

**(b)** $d = 1.1$ μm, $c = 0.1$ % v/v

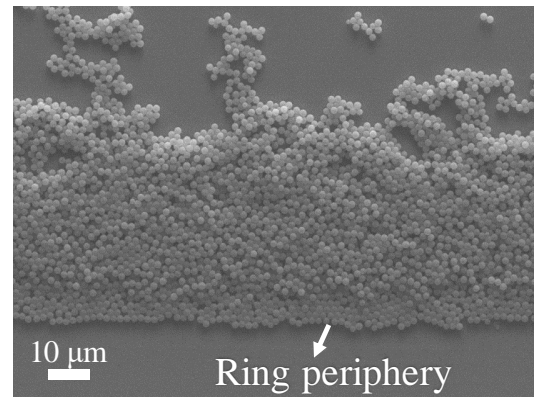

Figure 4. SEM of the surface of the ring for 0.1 μm (a) and 1.1 μm (b) particles. The concentration in both cases is 0.1 %.

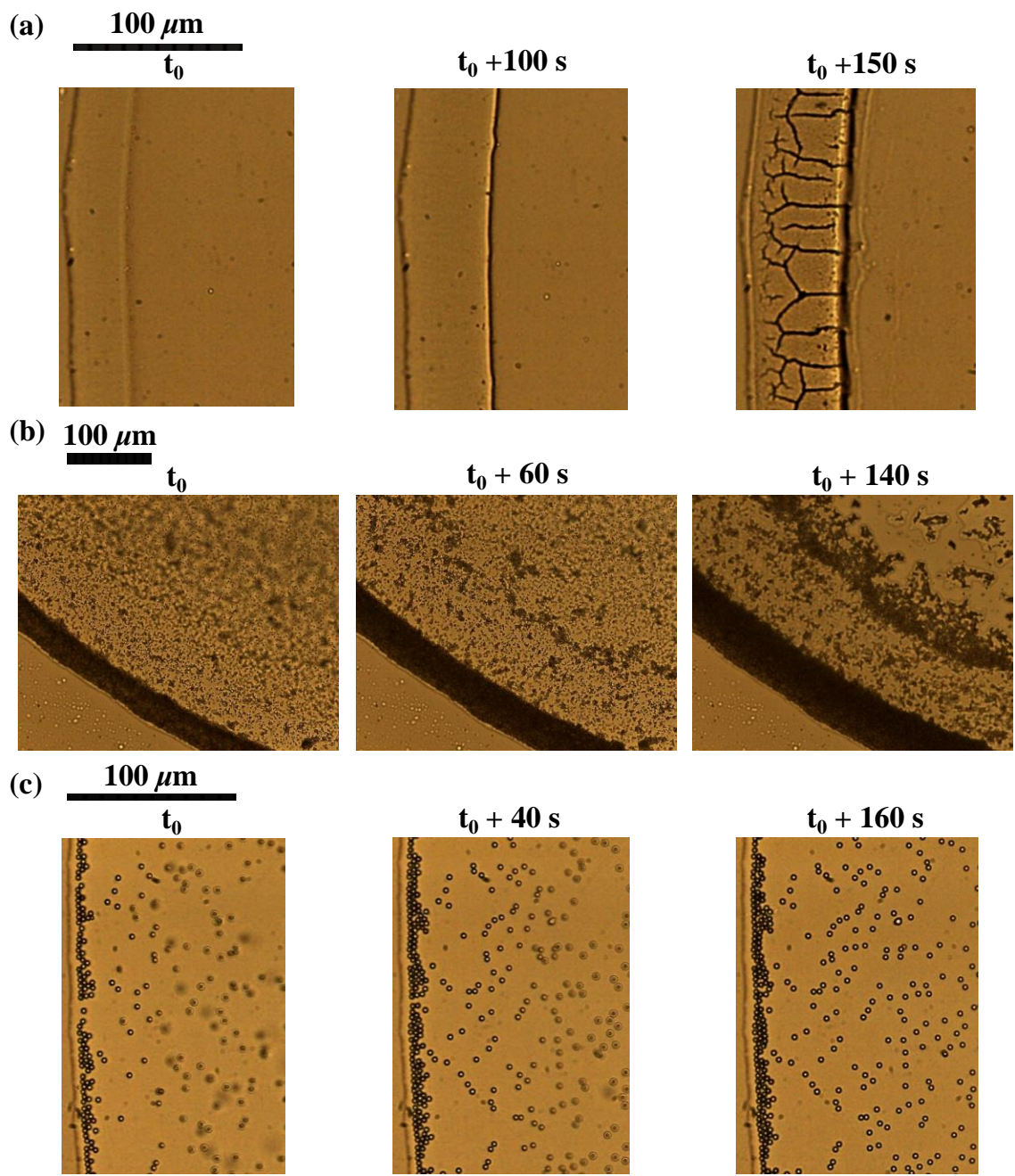

Figure 5. Time-sequence of images showing ring formation in late stages of the evaporation at $c = 0.1$ % for three cases of particles size (a) $d = 0.1$ μm (b) $d = 1.1$ μm (c) $d = 3.0$ μm. The ring periphery is on the left in each frame. Respective scale bar is shown on the top of each row. Associated movies SI.1, SI.2, and SI.3 are provided in the supporting information.

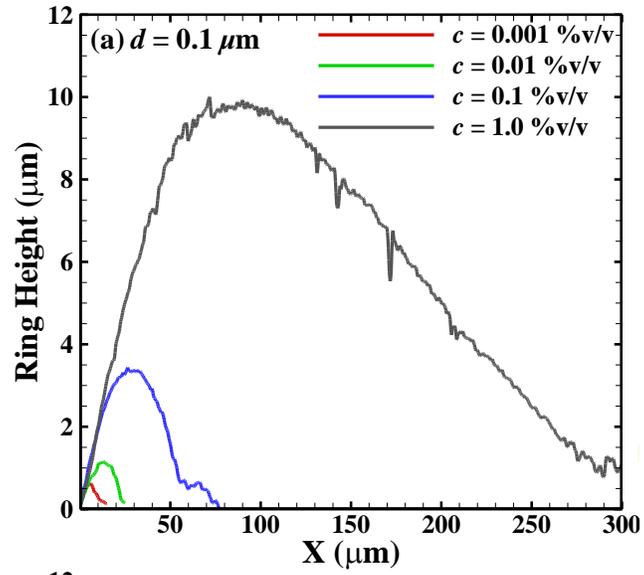
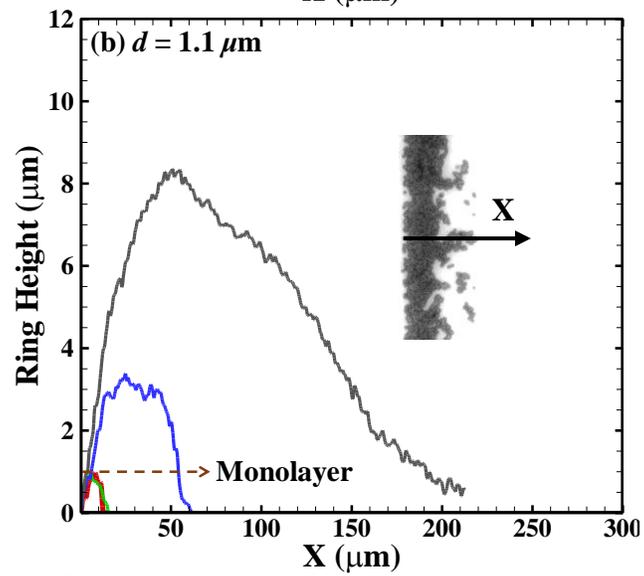
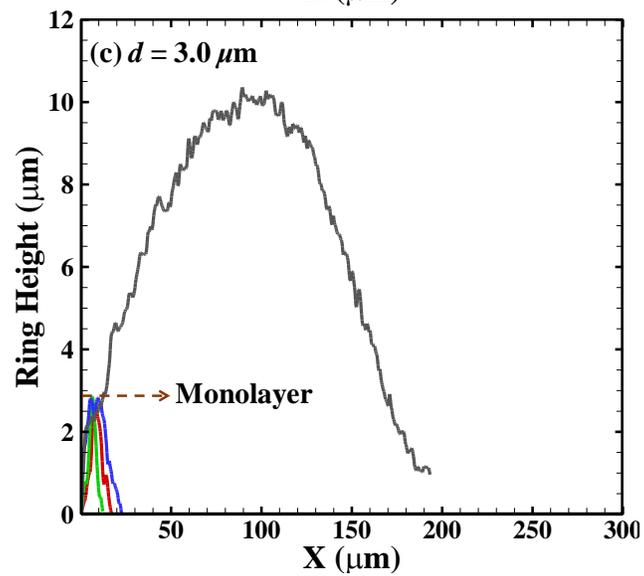

Figure 6: Average ring profiles measured using an optical profilometer and plotted for different cases of particles size (a) $d = 0.1$ $\mu$m, (b) $d = 1.1$ $\mu$m, and (c) $d = 3.0$ $\mu$m. For each case, different cases of concentrations, $c = 0.001, 0.01, 0.1$ and $1.0$ % are plotted. X represents the radial position in the deposit (shown as the inset in (b)) and $X = 0$ is the ring periphery.

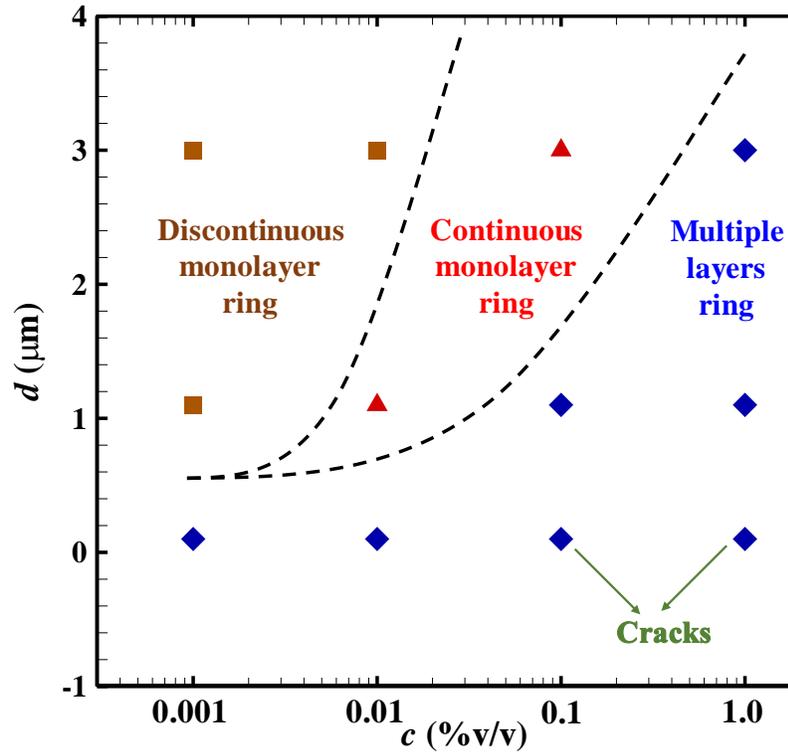

Figure 7: Regime map for classifying regimes of the discontinuous monolayer ring, continuous monolayer ring and multiple layers ring on particle concentration (c) - particle size (d) plane. Dashed lines serve as a guide to the eye to demarcate the regimes.

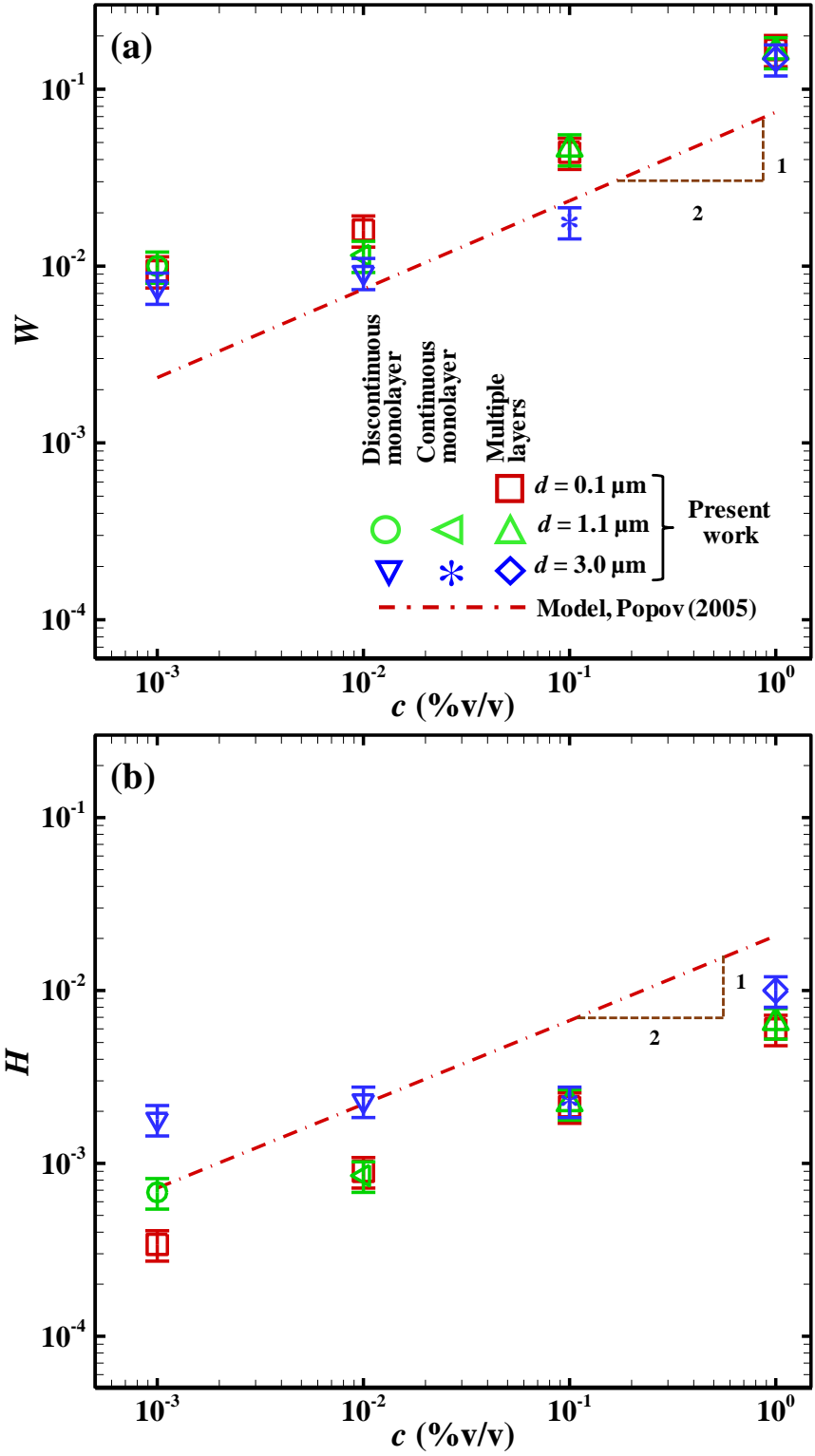

Figure 8: Comparison of measured ring dimensions with model predictions at different particles size, $d$ and particles concentration, $c$. (a) Non-dimensional ring width ($W$) (b) Non-dimensional ring height ($H$). Symbol and broken line represent measurement and model prediction, respectively.

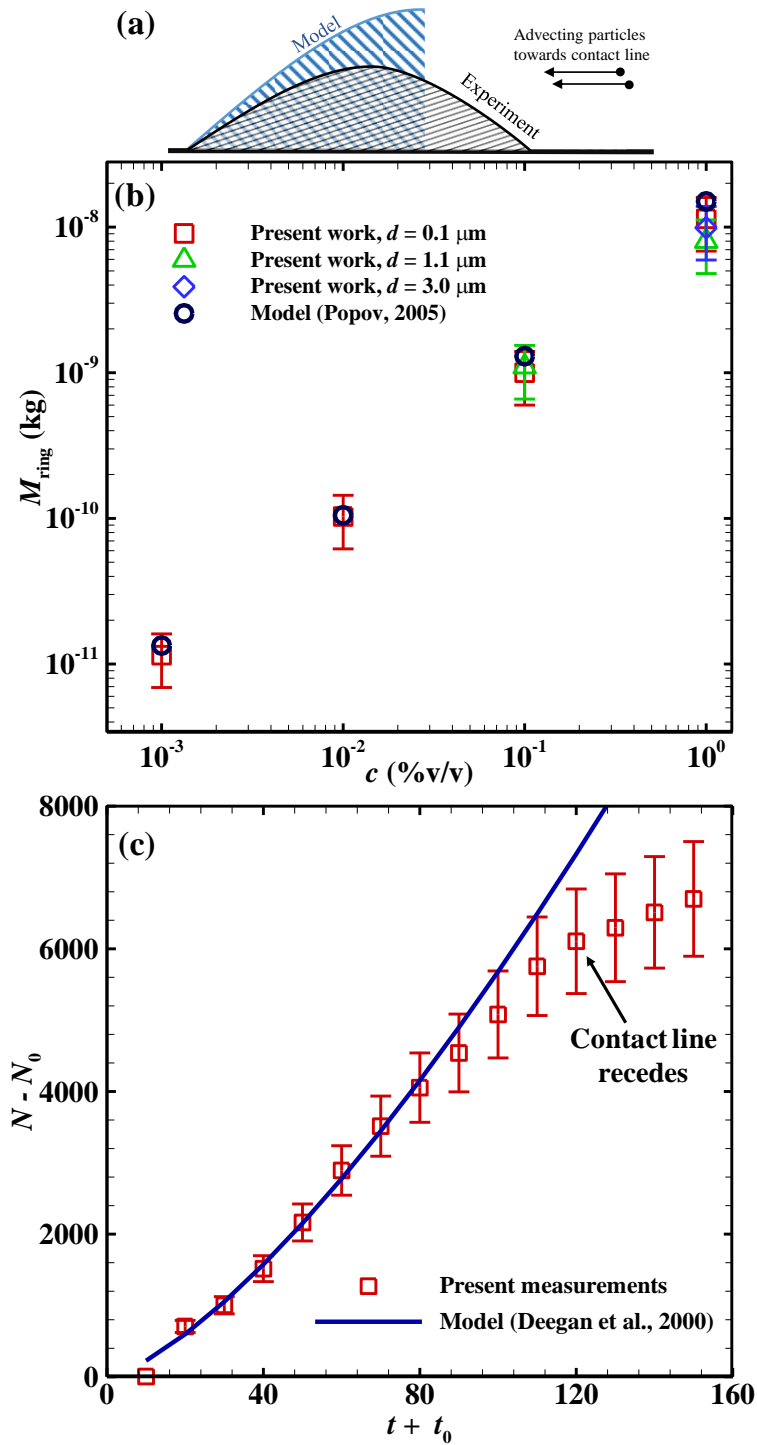

Figure 9. (a) Schematic showing a comparison between assumed ring-profile in the model and measured profile. (b) Comparison between the mass of the particles in the ring obtained by the measurements and the model at $d = 0.1$, $1.1$, and $3.0$ $\mu$m and different concentrations for multiple layers ring. (c) Count of particles advecting near the contact line as a function of time at $d = 3.0$ $\mu$m and $c = 0.1$ %.